# Research Data Explored: Citations versus Altmetrics

Isabella Peters[1], Peter Kraker[2], Elisabeth Lex[3], Christian Gumpenberger[4], and Juan Gorraiz[4]

[1] i.peters@zbw.eu
ZBW Leibniz Information Centre for Economics, Düsternbrooker Weg 120, D-24105 Kiel (Germany) & Kiel University, Christian-Albrechts-Platz 4, D-24118 Kiel (Germany)

[2] pkraker@know-center.at
Know-Center, Inffeldgasse 13, A-8010 Graz (Austria)

[3] elex@know-center.at
Graz University of Technology, Knowledge Technologies Institute, Inffeldgasse 13, A-8010 Graz (Austria)

[4] christian.gumpenberger | juan.gorraiz@univie.ac.at
University of Vienna, Vienna University Library, Dept of Bibliometrics, Boltzmanngasse 5, A-1090 Vienna (Austria)

**Abstract**
The study explores the citedness of research data, its distribution over time and how it is related to the availability of a DOI (Digital Object Identifier) in Thomson Reuters' DCI (Data Citation Index). We investigate if cited research data "impact" the (social) web, reflected by altmetrics scores, and if there is any relationship between the number of citations and the sum of altmetrics scores from various social media-platforms. Three tools are used to collect and compare altmetrics scores, i.e. PlumX, ImpactStory, and Altmetric.com. In terms of coverage, PlumX is the most helpful altmetrics tool. While research data remain mostly uncited (about 85%), there has been a growing trend in citing data sets published since 2007. Surprisingly, the percentage of the number of cited research data with a DOI in DCI has decreased in the last years. Only nine repositories account for research data with DOIs and two or more citations. The number of cited research data with altmetrics scores is even lower (4 to 9%) but shows a higher coverage of research data from the last decade. However, no correlation between the number of citations and the total number of altmetrics scores is observable. Certain data types (i.e. survey, aggregate data, and sequence data) are more often cited and receive higher altmetrics scores.

**Conference Topic**
Altmetrics - Citation and co-citation analysis

**Introduction**
Recently, data citations have gained momentum (Piwowar & Chapman, 2010; Borgman, 2012; Torres-Salinas, Martín-Martín, & Fuente-Gutiérrez, 2013). This is reflected, among others, in the development of data-level metrics (DLM), an initiative driven by PLOS, UC3 and DataONE[1], to track and measure activity on research data, and the recent announcement of CERN to provide DOIs for each dataset they share through their novel Open Data portal[2]. Data citations are citations included in the reference list of a publication that formally cite either the data that led to a research result or a data paper[3]. Thereby, data citations indicate the influence and reuse of data in scientific publications.

First studies on data citations showed that certain well-curated data sets receive far more citations or mentions in other articles than many traditional articles (Belter, 2014). Citations, however, are used as a proxy for the assessment of impact primarily in the "publish or perish" community; to consider other disciplines and stakeholders of research, such as industry, government and academia, and in a much broader sense, the society as a whole, altmetrics (i.e. social media-based indicators) are emerging as a useful instrument to assess the "societal" impact of research data or at least to provide a more complete picture of research uptake, besides more traditional usage and citation metrics (Bornman, 2014; Konkiel, 2013). Previous work on altmetrics for research data has mainly focused on motivations for data sharing, creating reliable data metrics and effective reward systems (Costas et al., 2012).

---

[1] http://escholarship.org/uc/item/9kf081vf
[2] https://www.datacite.org/news/cern-launches-data-sharing-portal.html
[3] http://www.asis.org/Bulletin/Jun-12/JunJul12_MayernikDataCitation.html

This study contributes to the research on data citations in describing their characteristics as well as their impact in terms of citations and altmetrics scores. Specifically, we tackle the following research questions:
- How often are research data cited? Which and how many of these have a DOI? From which repositories do research data originate?
- What are the characteristics of the most cited research data? Which data types and disciplines are the most cited? How does citedness evolve over time?
- To what extent are cited research data visible on various altmetrics channels? Are there any differences between the tools used for altmetrics scores aggregation?

**Data sources**
On the Web, a large number of data repositories are available to store and disseminate research data. The Thomson Reuters Data Citation Index (DCI), launched in 2012, provides an index of high-quality research data from various data repositories across disciplines and around the world. It enables search, exploration and bibliometric analysis of research data through a single point of access, i.e. the Web of Science (Torres-Salinas, Martín-Martín & Fuente- Gutiérrez, 2013). The selection criteria are mainly based on the reputation and characteristics of the repositories[4]. Three document types are available in the DCI: data set, data study, and repository. The document type "repository" can distort bibliometric analyses, because repositories are mainly considered as a source, but not as a document type.

First coverage and citation analyses of the DCI have been performed April-June 2013 by the EC3 bibliometrics group of Granada (Torres-Salinas, Jimenez-Contreras & Robinson-Garcia, 2014; Torres-Salinas, Robinson-Garcia & Cabezas-Clavijo, 2013). They found that data is highly skewed: Science areas accounted for almost 80% of records in the database and four repositories contained 75% of all the records in the database; 88% of all records remained uncited. In Science, Engineering and Technology citations are concentrated among datasets, whereas in the Social Sciences and Arts & Humanities, citations often refer to data studies.

Since these first analyses, DCI has been constantly growing, now indexing nearly two million records from high-quality repositories around the world. One of the most important enhancements of the DCI has undoubtedly been the inclusion of "figshare[5]" as new data source which led to an increase of almost a half million of data sets and 40.000 data studies (i.e. about one fourth of the total coverage in the database).

Gathering altmetrics data is quite laborious since they are spread over a variety of social media platforms which each offer different applications programming interfaces (APIs). Tools, which collect and aggregate these altmetrics data come in handy and are now fighting for market shares since also large publishers increasingly display altmetrics for articles (e.g.,

---
[4] http://thomsonreuters.com/data-citation-index, http://thomsonreuters.com/products/ip-science/04_037/dci-selection-essay.pdf
[5] http://figshare.com

Wiley[6]). There are currently three big altmetrics data providers: ImpactStory[7], Altmetric.com, and PlumX[8]. Whereas Altmetrics.com and PlumX focus more on gathering and providing data for institutions (e.g., publishers, libraries, or universities), ImpactStory's target group is the individual researcher who wants to include altmetrics information in her CV.

ImpactStory is a web-based tool, which works with individually assigned permanent identifiers (such as DOIs, URLs, PubMed IDs) or links to ORCID, Figshare, Publons, Slideshare, or Github to auto-import new research outputs like e.g. papers, data sets, slides. Altmetric scores from a large range of social media-platforms, including Twitter, Facebook, Mendeley, Figshare, Google+, and Wikipedia[9], can be downloaded as .json or .csv (as far as original data providers allow data sharing). With Altmetric.com, users can search within a variety of social media-platforms (e.g., Twitter, Facebook, Google+, or 8,000 blogs[10]) for keywords as well as for permanent identifiers (e.g., DOIs, arXiv IDs, RePEc identifiers, handles, or PubMed IDs). Queries can be restricted to certain dates, journals, publishers, social media-platforms, and Medline Subject Headings. The search results can be downloaded as .csv from the Altmetric Explorer (web-based application) or via the API. Plum Analytics or Plum X (the fee-based altmetrics dashboard) offers article-level metrics for so-called artifacts, which include articles, audios, videos, book chapters, or clinical trials[11]. Plum Analytics works with ORCID and other user IDs (e.g., from YouTube, Slideshare) as well as with DOIs, ISBNs, PubMed-IDs, patent numbers, and URLs. Because of its collaboration with EBSCO, Plum Analytics can provide statistics on the usage of articles and other artifacts (e.g., views to or downloads of html pages or pdfs), but also on, amongst others, Mendeley readers, GitHub forks, Facebook comments, and YouTube subscribers.

**Methodology**

We used DCI to retrieve the records of cited research data. All items published in the last 5.5 decades (1960-9, 1970-9, 1980-9, 1990-9, 2000-9, and 2010-4) with two or more citations (Sample 1, n=10,934 records) were downloaded and analysed. The criterion of having at least two citations is based on an operational reason (reduction of the number of items) as well as on a conceptual reason (to avoid self-citations). The following metadata fields were used in the analysis: available DOI or URL, document type, source, research area, publication year, data type, number of citations and ORCID availability[12]. The citedness in the database was computed for each decade considered in this study and investigated in detail for each year since 2000. We then analysed the distribution of document types, data types, sources and research areas with respect to the availability or non-availability of DOIs reported by DCI.

All research data with two or more citations and with an available DOI (n=2,907 items) were analysed with PlumX, ImpactStory, and Altmetric.com and their coverage on social media platforms and the altmetric scores was compared. All other items with 2 or more citations and an available URL (n=8,027) were also analysed in PlumX, the only tool enabling analyses based on URLs, and the results were compared with the ones obtained for items with a DOI.

---

[6] http://eu.wiley.com/WileyCDA/PressRelease/pressReleaseId-108763.html?campaign=wlytk-41414.4780439815
[7] https://impactstory.org
[8] https://plu.mx
[9] http://feedback.impactstory.org/knowledgebase/articles/367139-what-data-do-you-include-on-profiles
[10] http://support.altmetric.com/knowledgebase/articles/83335-which-data-sources-does-altmetric-track
[11] http://www.plumanalytics.com/metrics.html
[12] The DCI field "data type" was manually merged to more general categories; e.g. "survey data in social sciences" was merged with the category "survey data".

We also analysed the distribution of document types, data types, sources and research areas for all research data with 2 or more citations and at least one altmetric score (sample 2; n=301 items) with respect to the availability or non-availability of the permanent identifier DOI reported by DCI (items with DOI and URL or items with URL only).

Table 1. Results of DCI-based citation and altmetrics analyses for the last 5.5 decades.

| DCI | 1960-69 | 1970-79 | 1980-89 | 1990-99 | 2000-09 | 2010-14 |
|---|---|---|---|---|---|---|
| total # items | 6 040 | 23 712 | 43 620 | 186 965 | 2 096 023 | 1 627 668 |
| # items with > 2 citations | 5 | 110 | 360 | 956 | 4 727 | 4 777 |
| # items with at least 1 citation | 5 | 4207 | 7519 | 43749 | 239867 | 218440 |
| uncited % | 99.9% | 82.3% | 82.8% | 76.6% | 88.6% | 86.6% |
| items with DOI and >= 2 cit | 4 | 107 | 343 | 846 | 1381 | 226 |
| % with DOI and >=2 cit | 0.8 | 97.27% | 95.28% | 88.49% | 29.22% | 4.73% |
| with Altmetrics Data (PlumX) | 1 | 5 | 14 | 40 | 114 | 20 |
| % | 25.0% | 4.7% | 4.1% | 4.7% | 8.3% | 8.8% |
| items with URL only and >= 2 cit | 1 | 3 | 17 | 110 | 3 346 | 4551 |
| % with URL only and >=2 cit | 0.2 | 2.73% | 4.72% | 11.51% | 70.78% | 95.27% |
| with Altmetrics Data (PlumX) | 1 | 1 | 8 | 11 | 54 | 33 |
| % | 100.0% | 33.3% | 47.1% | 10.0% | 1.6% | 0.7% |

**Results and discussion**

**Part 1. General Results**

Table 1 gives an overview of the general results obtained in this study. The analysis revealed a high uncitedness of research data, which corresponds to the findings of Torres-Salinas, Martin-Martin and Fuente-Gutiérrez (2013). A more detailed analysis for each year (see Table 2) shows, however, that the citedness is comparatively higher for research data published in recent years (published after 2007) although the citation window is shorter.

Table 2. Evolution of uncitedness in DCI in the last 14 years.

| PY | Items | uncited | % uncited |
|---|---|---|---|
| 2000 | 28282 | 18152 | 64.18% |
| 2001 | 36397 | 25367 | 69.70% |
| 2002 | 64781 | 51464 | 79.44% |
| 2003 | 115997 | 93538 | 80.64% |
| 2004 | 141065 | 122802 | 87.05% |
| 2005 | 212781 | 178146 | 83.72% |
| 2006 | 299443 | 275216 | 91.91% |
| 2007 | 362405 | 333136 | 91.92% |
| 2008 | 398931 | 364236 | 91.30% |
| 2009 | 435941 | 394099 | 90.40% |
| 2010 | 390957 | 349623 | 89.43% |
| 2011 | 270932 | 224790 | 82.97% |
| 2012 | 492534 | 428752 | 87.05% |
| 2013 | 448489 | 386507 | 86.18% |
| 2014 | 24756 | 19556 | 78.99% |

**Table 3. Overview on citation distribution of Sample 1 (n=10,934 items).**

| items with at least 2 citations | Document Type | # items | Total Citations | Mean Citations | Maximum Citations | Standard Deviation | Variance |
|---|---|---|---|---|---|---|---|
| all | Data set | 5641 | 17984 | 3.19 | 121 | 3.38 | 11.46 |
| | Data study | 5242 | 91623 | 17.48 | 1236 | 50.22 | 2521.67 |
| | Repository | 51 | 10076 | 197.57 | 3193 | 618.73 | 382824.45 |
| | Total | 10934 | 119683 | 10.95 | 3193 | 56.39 | 3179.49 |
| with DOI | Data set | 342 | 977 | 2.86 | 52 | 3.86 | 14.93 |
| | Data study | 2565 | 53293 | 20.78 | 1236 | 63.44 | 4024.45 |
| | Total | 2907 | 54270 | 18.67 | 1236 | 59.88 | 3585.92 |
| with URL only | Data set | 5299 | 17007 | 3.21 | 121 | 3.35 | 11.23 |
| | Data study | 2677 | 38330 | 14.32 | 272 | 32.59 | 1062.31 |
| | Repository | 51 | 10076 | 197.57 | 3193 | 618.73 | 382824.45 |
| | Total | 8027 | 65413 | 8.15 | 3193 | 54.80 | 3003.30 |

The results also show a very low percentage of altmetrics scores available for research data with two or more citations (see Table 1). But, two different trends can be observed: the percentage of data with DOI referred to on social media-platforms is steadily increasing while the percentage of data with URL only is steadily decreasing in the same time frame.

The percentage of research data with altmetrics scores in PlumX, the tool with the highest average in this study, is lower than expected (ranging between 4 and 9%) but actually has doubled for data published in the last decades, which confirms the interest in younger research data and an increase in social media activity of the scientific community in recent years.

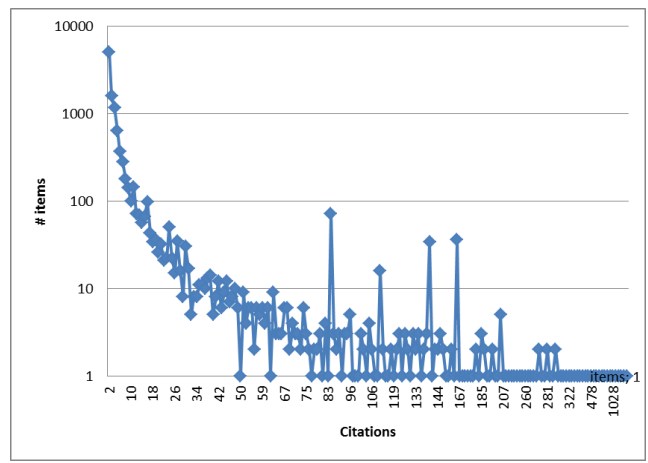

**Figure 1. Citation distribution of Sample 1 (logarithmic scale).**

## Part 2: Results for Sample 1

Table 3 shows the citation distribution of Sample 1 (10,934 items with at least two citations in DCI) for items with DOI or URL only separated according to the three main DCI document types (data set, data study, and repository[13]). The results reveal that almost half of the data studies have a DOI (48.9%) but only few data sets do so. Data studies are on average more often cited than data sets (17.5 vs. 3.2 citations per item), and data studies with a DOI attract more citations (mean values) than those with a URL (20 vs. 14 citations per item).

---

[13] Table 3 includes repositories as document type to illustrate the citation volume in DCI.

Table 4. Analysis of Sample 1 by sources (repositories) (n=10,934 items).

| Sources (with DOI) | # items | # citations | Sources (with URL) | # items | # citations |
|---|---|---|---|---|---|
| Inter-university Consortium for Political and Social Research | 2530 | 53041 | miRBase | 3456 | 10209 |
| Worldwide Protein Data Bank | 229 | 458 | Cancer Models Database | 864 | 2698 |
| Oak Ridge National Laboratory Distributed Active Archive Center for Biogeochemical Dynamics | 108 | 508 | UK Data Archive | 836 | 25479 |
| Archaeology Data Service | 21 | 75 | European Nucleotide Archive | 361 | 1346 |
| 3TU.Datacentrum | 8 | 22 | Gene Expression Omnibus | 353 | 754 |
| SHARE - Survey of Health, Ageing and Retirement in Europe | 4 | 151 | National Snow & Ice Data Center | 298 | 2796 |
| World Agroforestry Centre | 3 | 6 | Australian Data Archive | 264 | 2469 |
| Dryad | 2 | 4 | Australian Antarctic Data Centre | 249 | 1621 |
| GigaDB | 2 | 5 | nmrshiftdb2 | 219 | 445 |
| | | | Finnish Social Science Data Archive | 183 | 913 |

There were only few repositories (51) in the data set; it is the document type, which attracts the most citations per item. This finding is in line with the results of Belter (2014) who also found aggregated data sets – Belter calls them "global-level data sets" – to be more cited. However, such citing behaviour has a negative side effect on repository content (i.e., the single data sets) since it is not properly attributed in favour of citing the repository as a whole.

The high values of the standard deviation and variance illustrate the skewness of the citation distribution (see Figure 1). Almost half of the research data (4,974 items; 45.5%) have only two citations. Six items, two repositories and four data studies, from different decades (PY=1981, 1984, 1995, 2002, 2011, and 1998, sorted by descending number of citations) had more than 1,000 citations and account for almost 30% of the total number of citations.

Table 4 shows the top 10 repositories by the number of items. Considering the number of citations, there are three other repositories which account for more than 1,000 citations each: Manitoba Centre for Health Policy Population Health Research Data Repository (29 items; 1,631 citations), CHILDES - Child Language Data Exchange System (1 item; 3,082 citations), and World Values Survey (1 item; 3,193 citations). Interestingly, although "figshare" accounts for almost 25% of the DCI, no item from "figshare" was cited at least twice in DCI. We also noted that the categorization of "figshare" items is missing. All items are assigned to the Web of Science category (WC) "Multidisciplinary Sciences" or the Research Area (SU) "Science & Technology/Other Topics" preventing detailed topic-based citation analyses. Furthermore, only nine items from Sample 1 were related to an ORCID, three data sets with a DOI, and three data sets and data studies with a URL.

Considering their origin, considerable differences were reported in Sample 1 for items with or without a DOI (see Table 4). All twice or more frequently cited research data with a DOI are archived in nine repositories, while 92 repositories account for research data without a DOI.

**Table 5. Analysis of Sample 1 by data types (manually merged), top 10 types (n=10,934 items).**

| Data Types (with DOI) | # items | # citations | Data Types (with URL only) | # items | # citations |
|---|---|---|---|---|---|
| survey data | 1734 | 43686 | sequence data | 3408 | 10458 |
| administrative records data | 302 | 3326 | profiling by array, gen, etc | 352 | 752 |
| aggregate data | 274 | 9440 | Individual (micro) level | 240 | 9024 |
| event/transaction data | 210 | 2400 | Numeric data | 216 | 4317 |
| clinical data | 118 | 3469 | Structured questionnaire | 155 | 673 |
| census/enumeration data | 109 | 1019 | survey data | 127 | 1315 |
| protein structure | 95 | 190 | Seismic:Reflection:MCS | 47 | 185 |
| observational data | 30 | 575 | statistical data | 41 | 1352 |
| program source code | 10 | 116 | Digital media | 40 | 290 |
| roll call voting data | 8 | 236 | EXCEL | 25 | 101 |

**Table 6. Sample 1 by research areas and document types, top 10 areas (n=10,934 items).**

| | with DOI | | | | | with URL only | | | |
|---|---|---|---|---|---|---|---|---|---|
| | # Items | | # citations | | | # Items | | # citations | |
| Research Area | Data set | Data study | Data set | Data study | Research Area | Data set | Data study | Data set | Data study |
| Criminology & Penology | | 471 | | 4403 | Genetics & Heredity | 4658 | 159 | 14024 | 571 |
| Sociology | | 432 | | 7930 | Meteorology & Atmospheric Sciences | 91 | 298 | 493 | 2796 |
| Government & Law | | 352 | | 10399 | Biochemistry & Molecular Biology; Genetics & Heredity | | 353 | | 754 |
| Demography | | 317 | | 9178 | Sociology | | 286 | | 1994 |
| Health Care Sciences & Services | | 290 | | 8170 | Physics | 5 | 214 | 10 | 435 |
| Biochemistry & Molecular Biology | 229 | | 458 | | Business & Economics; Sociology | | 143 | | 12665 |
| Business & Economics | | 204 | | 3083 | Biochemistry & Molecular Biology; Spectroscopy | 129 | | 383 | |
| Environmental Sciences & Ecology; Geology | 108 | | 508 | | Oceanography; Geology | 114 | | 353 | |
| Education & Educational Research | | 69 | | 1881 | Demography; Sociology | | 103 | | 5673 |
| Family Studies | | 68 | | 2268 | Sociology; Demography; Communication | | 84 | | 393 |

Table 5 shows that there are big differences between the most cited data types when considering research data with a DOI or with a URL. Survey data, aggregate data, and clinical data are the most cited ones of the first group (with a DOI), while sequence data and numerical and individual level data are the most cited data types of the second group (with a URL). Apart from survey data, there is no overlap in the top 10 data types indexed in DCI. Similar results were obtained when considering data sets and data studies separately.

Disciplinary differences become apparent in the citations of DOIs and URLs as well as in the use of certain document types. As shown in Table 6 it is more common to refer to data studies via DOIs in the Social Sciences than in the Natural and Life Sciences, where the use of URLs for both data studies and data sets is more popular. Torres-Salinas, Jimenez-Contreras and Robinson-Garcia (2014) also report that citations in Science, Engineering and Technology citations are concentrated on data sets, whereas the majority of citations in the Social Sciences and Arts & Humanities refer to data studies. Table 6 suggests that these differences could be related to the availability of a DOI.

**Table 7. Citation and altmetrics results of Sample 2 (n=301 items) according to document type.**
*8 items with URL found in PlumX could not properly be identified (broken URL, wrong item, etc.)

|  | *Document Type* | *# items* | *Total Citations* | *Mean Citations* | *Maximum Citations* | *Standard Deviation* | *Variance* |
|---|---|---|---|---|---|---|---|
| **with DOI** | Data set | 15 | 173 | 11.53 | 52 | 13.75 | 189.12 |
|  | Data study | 179 | 6716 | 37.52 | 1135 | 107.36 | 11525.43 |
|  | Total | 194 | 6889 | 35.51 | 1135 | 103.40 | 10691.82 |
|  | *Document Type* | *# items* | *Total Scores* | *Mean Scores* | *Maximum Scores* | *Standard Deviation* | *Variance* |
|  | Data set | 15 | 34 | 2.27 | 6 | 1.75 | 3.07 |
|  | Data study | 179 | 710 | 3.97 | 64 | 7.42 | 55.09 |
|  | Total | 194 | 752 | 376.00 | 748 | 526.09 | 276768.00 |
| **with URL only** | *Document Type* | *# items* | *Total Citations* | *Mean Citations* | *Maximum Citations* | *Standard Deviation* | *Variance* |
|  | Data set | 24 | 172 | 7.17 | 46 | 10.12 | 102.41 |
|  | Data study | 31 | 779 | 25.13 | 272 | 51.67 | 2669.65 |
|  | Repository | 44 | 9677 | 219.93 | 3193 | 662.92 | 439464.20 |
|  | Total* | 99 | 10628 | 107.35 | 3193 | 451.61 | 203954.50 |
|  | *Document Type* | *# items* | *Total Scores* | *Mean Scores* | *Maximum Scores* | *Standard Deviation* | *Variance* |
|  | Data set | 24 | 428 | 17.83 | 378 | 76.75 | 5890.23 |
|  | Data study | 31 | 664 | 21.42 | 213 | 53.25 | 2835.65 |
|  | Repository | 44 | 3961 | 90.02 | 1150 | 198.53 | 39415.70 |
|  | Total* | 99 | 5319 | 49.71 | 1150 | 139.82 | 19549.38 |

**Part 3: Results for Sample 2**

Sample 2 comprises all items from DCI satisfying the following criteria: two or more citations in DCI, a DOI or a URL and at least one altmetrics score in PlumX (n=301 items). Table 7 shows the general results for this sample. The total number of altmetrics scores is lower than the number of citations for all document types with or without a DOI. Furthermore, the mean altmetrics score is higher for data studies than for data sets.

Tables 8 and 9 show the distributions of data types and subject areas in this sample. Most data with DOI are survey data, aggregate data, event over transaction data, whereas sequence data and images are most often referred to via URL only (see Table 8). Microdata with DOI and spectra with URL only are the data types with the highest altmetrics scores per item. Concerning subject areas the results of Table 9 are very similar to the results of Table 6. Given the small sample size it is, however, notable that in some subject areas, e.g. Archaeology, research data receive more interest in social media (i.e. altmetrics scores), than via citations in traditional publications. This is confirmed by the missing correlation between citations and altmetrics scores for this sample (see Figure 2). Both cases clearly demonstrate that altmetrics can complement traditional impact evaluation. Nevertheless, coverage of research data in social media is still low, e.g. from the nine repositories whose data studies and data sets were cited twice in DCI and had a DOI (see Table 4), only five items had altmetrics scores in PlumX, and only one DOI item of Sample 2 included an ORCID.

**Table 8. Citation and altmetrics overview of Sample 2 (n=301 items) according to their data type**
(Field DY; no aggregated counts, "document type" "repository" (34 items) not included.

| Data Type (with DOI) | # items | total citations | mean citations | total scores | mean scores | Data Type (with URL only) * | # items | total citations | mean citations | total scores | mean scores |
|---|---|---|---|---|---|---|---|---|---|---|---|
| survey data | 110 | 5276 | 47.96 | 353 | 3.21 | miRNA sequence data | 15 | 71 | 4.73 | 21 | 1.40 |
| aggregate data | 26 | 793 | 30.50 | 80 | 3.08 | FITS images; spectra; calibrations; redshifts | 4 | 248 | 62 | 16 | 4.00 |
| event/transaction data | 19 | 414 | 21.79 | 43 | 2.26 | statistical data | 3 | 333 | 111 | 22 | 7.33 |
| administrative records data | 13 | 125 | 9.62 | 58 | 4.46 | Expression profiling by array | 3 | 6 | 2 | 4 | 1.33 |
| clinical data | 11 | 314 | 28.55 | 26 | 2.36 | Sensor data; survey data | 2 | 51 | 25.5 | 10 | 5.00 |
| census/enumeration data | 8 | 90 | 11.25 | 14 | 1.75 | Quantitative | 2 | 35 | 17.5 | 10 | 5.00 |
| observational data | 4 | 99 | 24.75 | 7 | 1.75 | images | 1 | 20 | 20 | 3 | 3.00 |
| Longitudinal data; Panel Data; Micro data | 2 | 79 | 39.50 | 46 | 23.00 | images; spectra | 1 | 4 | 4 | 102 | 102.00 |
| roll call voting data | 2 | 178 | 89.00 | 3 | 1.50 | table | 1 | 9 | 9 | 1 | 1.00 |
| machine-readable text | 1 | 5 | 5.00 | 1 | 1.00 | redshifts; spectra | 1 | 5 | 5 | 213 | 213.00 |
| program source code | 1 | 2 | 2.00 | 1 | 1.00 | images; spectra; astrometry | 1 | 2 | 2 | 90 | 90.00 |

**Table 9. Citation and altmetrics overview of Sample 2 according to their subject area.**

| with DOI | | | | with URL only | | | |
|---|---|---|---|---|---|---|---|
| Subject Areas | # items | # citations | # scores | Subject Areas | # items | # citations | # scores |
| Sociology | 35 | 1226 | 213 | Genetics & Heredity | 26 | 492 | 654 |
| Government & Law | 28 | 793 | 53 | Meteorology & Atmospheric Sciences | 15 | 166 | 28 |
| Criminology & Penology | 22 | 317 | 42 | Astronomy & Astrophysics | 9 | 933 | 427 |
| Health Care Sciences & Services | 14 | 1498 | 70 | Biochemistry & Molecular Biology; Genetics & Heredity | 5 | 22 | 557 |
| Environmental Sciences & Ecology; Geology | 14 | 171 | 33 | Cell Biology | 4 | 13 | 383 |
| Demography | 12 | 433 | 28 | Health Care Sciences & Services; Business & Economics | 3 | 335 | 68 |
| Family Studies | 10 | 166 | 26 | Genetics & Heredity; Biochemistry & Molecular Biology | 2 | 27 | 36 |
| Archaeology | 10 | 47 | 139 | Business & Economics | 2 | 35 | 10 |
| Education & Educational Research | 9 | 661 | 40 | Health Care Sciences & Services | 2 | 423 | 2 |
| International Relations | 9 | 384 | 46 | Communication; Sociology; Telecommunications | 2 | 51 | 10 |

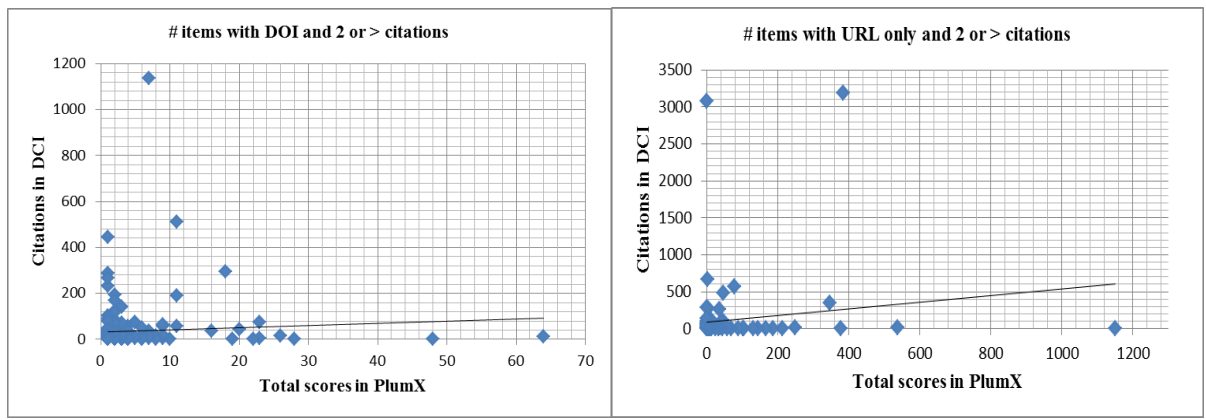

**Figure 2. Citations DCI versus scores in PlumX for items with (left) and without (right).**

**Part 4. Selected altmetrics scores and comparison of the results of three altmetrics tools**

Table 10 shows the general results obtained in PlumX according to PlumX's aggregation groups (i.e., captures, social media, mentions, and usage) for all document types and with or without DOI. While DOIs for data sets seem to be important in order to get captures (mainly in Mendeley), a URL is sufficient for an inclusion in social media tools like Facebook, Twitter, etc.

The top 10 research data-DOIs attracting two or more citations and with at least one entry in PlumX are shown in Table 11. We can observe that cited research data attracts more citations than altmetrics scores, and that there is no correlation between highly cited and highly scored research data.

The comparison of altmetrics aggregation tools also revealed that ImpactStory only found Mendeley reader statistics for the research data: 78 DOIs had 257 readers. Additionally, ImpactStory found one other DOI in Wikipedia. ImpactStory found five items, which have not been found by PlumX, although they all solely relied on Mendeley Data. The Mendeley data scores were exactly the same in PlumX and in ImpactStory. On the other hand, PlumX found 18 items that were not available via ImpactStory. These research data were distributed on social media platforms (mostly shares in Facebook) and one entry has been used via click on a Bitly-URL (Usage:Clicks:Bitly).The tool Altmetric.com found only one from 194 items.

As already reported in Jobmann et al. (2014), PlumX is the tool with the highest coverage of research products found on social media-platforms. Whereas Mendeley is well covered in ImpactStory, no other altmetrics score were found for the data set used in this study.

**General Conclusions**

Most of the research data still remain uncited (approx. 86%) and total altmetrics scores found via aggregation tools are even lower than the number of citations. However, research data published from 2007 onwards have gradually attracted more citations reflecting a bias towards more recent research data. No correlation between citation and altmetrics scores could be observed in a preliminary analysis: neither the most cited research data nor the most cited sources (repositories) received the highest scores in PlumX.

In the DCI, the availability of cited research data with a DOI is rather low. A reason for this may be the increase of available research data in recent years. Furthermore, the percentage of cited research data with a DOI has not increased as expected, which indicates that citations do not depend on this standard identifier in order to be processed by the DCI.

**Table 10. PlumX altmetrics scores for all document types with or without DOI.**

| | | with DOI | | | with URL only | | | |
|---|---|---|---|---|---|---|---|---|
| | Document Type | Data set | Data study | Total | Data set | Data study | Repository | Total |
| | *# items* | 15 | 179 | **194** | 24 | 31 | 44 | **99** |
| *Captures* | *Sum* | 32 | 471 | **503** | 0 | 0 | 30 | **30** |
| | *Mean* | 2.13 | 2.63 | **2.59** | 0.00 | 0.00 | 0.68 | **0.28** |
| | *Max* | 6 | 48 | **48** | 0 | 0 | 23 | **23** |
| *Social Media* | *Sum* | 1 | 220 | **221** | 407 | 281 | 3060 | **3890** |
| | *Mean* | 0.07 | 1.23 | **1.14** | 16.96 | 9.06 | 69.55 | **36.36** |
| | *Max* | 1 | 58 | **58** | 366 | 119 | 1008 | **1008** |
| *Mentions* | *Sum* | 1 | 13 | **14** | 13 | 62 | 433 | **629** |
| | *Mean* | 0.07 | 0.07 | **0.07** | 0.54 | 2.00 | 9.84 | **5.88** |
| | *Max* | 1 | 4 | **4** | 12 | 31 | 119 | **120** |
| *Usage* | *Sum* | 0 | 6 | **6** | 8 | 321 | 438 | **770** |
| | *Mean* | 0.00 | 0.03 | **0.03** | 0.33 | 10.35 | 9.95 | **7.20** |
| | *Max* | 0 | 6 | **6** | 4 | 187 | 92 | **187** |
| *Total entries* | | 34 | 710 | **744** | 428 | 664 | 3961 | **5319** |
| *% Captures* | | 94.1% | 66.3% | **67.6%** | 0.0% | 0.0% | 0.8% | **0.6%** |
| *% Social Media* | | 2.9% | 31.0% | **29.7%** | 95.1% | 42.3% | 77.3% | **73.1%** |
| *% Mentions* | | 2.9% | 1.8% | **1.9%** | 3.0% | 9.3% | 10.9% | **11.8%** |
| *% Usage* | | 0.0% | 0.8% | **0.8%** | 1.9% | 48.3% | 11.1% | **14.5%** |

Nevertheless, data studies with a DOI attract more citations than those with a URL. Despite the low number of research data with a DOI in general, surprisingly, the DOI in cited research data has so far been more embraced in the Social Sciences than in the Natural Sciences.

Furthermore, our study shows an extremely low number of research data with two or more citations (only nine out of around 10,000) related to an ORCID. Only three of them had a DOI likewise. This illustrates that we are still a far cry from the establishment of permanent identifiers and their optimal interconnectedness in a data source.

The low percentage of altmetrics scores for research data with two or more citations corroborates a threefold hypothesis: First, research data are either rarely published or not findable on social media-platforms, because DOIs or URLs are not used in references thus resulting in a low coverage of items. Second, research data are not widely shared on social media by the scientific community so far which would result in higher altmetrics scores[14]. Third, the reliability of altmetrics aggregation tools is questionable as the results on the coverage of research data on social media-platforms differ widely between tools. However, the steadily increasing percentage of cited research data with DOI suggests that the adoption of this permanent identifier increases the online visibility of research data and inclusion in altmetrics tools (since they heavily rely on DOIs or other permanent identifiers for search).

A limitation of our study is that the results rely on the indexing quality of the DCI. Our analysis shows that the categorisation in DCI is problematic at times. This is illustrated by the fact that all items from figshare, which is one of the top providers of records, are categorised into "Miscellaneous". Also, the category "repository" is rather a source than a document type. Such incorrect assignments of data types and disciplines can easily lead to wrong interpretations in citation analyses. Furthermore, it should be taken into account that citation counts are not always traceable.

---

[14] figshare lately announced a deal with Altmetric.com which might increase the visibility of altmetrics with respect to data sharing: http://figshare.com/blog/The_figshare_top_10_of_2014_according_to_altmetric/142

Table 11. Top 10 Research Data with DOI according to the total scores in PlumX.

| DOI | SO | PY | Captures:Readers:Mendeley | Social Media:+1s:Google+ | Social Media:Shares:Facebook | Social Media:Likes:Facebook | Social Media:Tweets:Twitter | Mentions:Comments:Facebook | # total Scores | # Citations |
|---|---|---|---|---|---|---|---|---|---|---|
| 10.5284/1000415 | ADS | 2012 | 2 | | 13 | 45 | | 4 | 64 | 13 |
| 10.3886/icpsr13580 | IUC | 2005 | 48 | | | | | | 48 | 3 |
| 10.5284/1000397 | ADS | 2011 | | | 14 | 12 | | 2 | 28 | 2 |
| 10.3886/icpsr06389 | IUC | 2007 | 25 | 1 | | | | | 26 | 14 |
| 10.6103/share.w4.111 | SHARE | 2004 | | | 8 | 15 | | | 23 | 74 |
| 10.6103/share.w4.111 | SHARE | 2010 | | | 8 | 15 | | | 23 | 5 |
| 10.3886/icpsr13611 | IUC | 2006 | 22 | | | | | | 22 | 3 |
| 10.3886/icpsr02766 | IUC | 2007 | 20 | | | | | | 20 | 44 |
| 10.5284/1000381 | ADS | 2009 | | 2 | 3 | 10 | 3 | 1 | 19 | 2 |
| 10.3886/icpsr09905 | IUC | 1994 | 18 | | | | | | 18 | 295 |
| 10.3886/icpsr08624 | IUC | 2010 | 16 | | | | | | 16 | 36 |
| 10.3886/icpsr04697 | IUC | 2009 | 11 | | | | | | 11 | 510 |
| 10.3886/icpsr06716 | IUC | 2007 | 11 | | | | | | 11 | 59 |
| 10.3886/icpsr20240 | IUC | 2008 | 11 | | | | | | 11 | 190 |
| 10.3886/icpsr20440 | IUC | 2007 | 3 | | | | 7 | | 10 | 3 |

Finally, citations of research data should be studied in more detail. They certainly differ from citations of papers relying on these data with regard to dimension and purpose. For example, we found that entire repositories are proportionally more often cited than single data sets, which was confirmed by a former study (Belter, 2014). Therefore, it will be important to study single repositories (such as figshare) in more detail. It is crucial to further explore the real meaning and rationale of research data citations and how they depend on the nature and structure of the underlying research data, e.g., in terms of data curation and awarding of DOIs. Also, little is known about how data citations complement and differ from data sharing and data usage activities as well as altmetrics.

## Acknowledgments

This analysis was done within the scope of e-Infrastructures Austria (http://e-infrastructures.at/). The authors thank Dr. Uwe Wendland (Thomson Reuters) and Stephan Buettgen (EBSCO) for granted trial access to Data Citation Index resp. PlumX. The Know-Center is funded within the Austrian COMET program – Competence Centers for Excellent Technologies - under the auspices of the Austrian Federal Ministry of Transport, Innovation and Technology, the Austrian Federal Ministry of Economy, Family and Youth, and the State of Styria. COMET is managed by the Austrian Research Promotion Agency FFG.